\documentclass[aps,prd,amsmath,amsfonts,amssymb,eqsecnum,nofootinbib,
  floatfix,secnumarabic,preprintnumbers,superscriptaddress,nobalancelastpage,
  onecolumn,notitlepage]{revtex4}
\usepackage{color,graphicx,hyperref,dsfont,slashed,multirow}
\usepackage{placeins}
\usepackage{longtable}

\hypersetup{bookmarks=true,unicode=true,pdftoolbar=true,pdfmenubar=true,
  pdffitwindow=false,pdfstartview={FitH},pdftitle={mBLRISS},
  pdfauthor={~A.~Kuday,F.~\"{O}zok, ~U. ~Saka}, pdfsubject={Subject},
  pdfcreator={Creator},pdfproducer={Producer}, pdfkeywords={dark matter}{Effective Field Theory}{Relic Density}{Relic Abundance}{Monojet}{Dijet}{Missing Transverse Energy}
  pdfnewwindow=true,colorlinks=true,
  linkcolor=blue,citecolor=magenta,filecolor=magenta,urlcolor=cyan}
\newcommand{\be}{\begin{equation}}
\newcommand{\ee}{\end{equation}}
\def\bsp#1\esp{\begin{split}#1\end{split}}
\renewcommand{\figureautorefname}{Fig.}

\def\sectionautorefname~#1\null{Sec.~(#1)\null}
\def\subsectionautorefname~#1\null{sub--Sec.~(#1)\null}
\def\figureautorefname~#1\null{Fig.~#1\null}
\def\tableautorefname~#1\null{Table~#1\null}
\def\equationautorefname~#1\null{Eq.~#1\null}

\begin{document}
\date{\today}
\preprint{-}
\title{Analysis of Direct and Indirect Detection of Fermionic Dark Matter of 6-Dimensional Effective Field Theory}
\author{Ay\c{s}e  El\c{c}{\.i}bo\u{g}a  Kuday}
\email{ayse.kuday@msgsu.edu.tr}
\affiliation{Department of Physics, Mimar Sinan Fine Arts University,  34380, {\.I}stanbul, TURKEY}

\author{Ferhat \"Ozok}
\email{Ferhat.Ozok@cern.ch}
\affiliation{Department of Physics, Mimar Sinan Fine Arts University,  34380, {\.I}stanbul, TURKEY}

\author{Erdin\c{c} Ula\c{s} Saka}
\email{ulassaka@istanbul.edu.tr}
\affiliation{{\.I}stanbul University Faculty of Science, Department of Physics, 34134 , {\.I}stanbul, TURKEY}

\vspace{10pt}
\begin{abstract}

We present an analysis of fermionic dark matter (DM) in the context of 6 dimensional Effective Field Theory (EFT). We also compared the result generated via 
the 6-dimensional EFT analysis with the current experimental results for dark matter searches. These experiments are  methodically categorised as direct and indirect search 
and present some constraints on dark matter model parameters of 6-dimensional EFT. 
We constructed a new set of tools ensuring DM researches in various platforms. The model parameters are presented to guide DM production in colliders by taking account of the upper limits at direct and indirect searches. In this paper we apply our approach for fermionic case to test the verification of the method.
There are various type of search methods for DM, each depends on type of interaction of dark matter with SM particles. Finally we analysed fermionic DM candidate of 6-dimensional Effective Field Theory (EFT) at the platforms of DM searches. A new set of numerical tools is specified for 6-dimensional fermionic DM model, and these tools are also tested. 

\end{abstract}


\maketitle

\section{Introduction}\label{sec:intro}
Dark matter is a theoretical substance that makes up $85\%$ of the Universe's mass. On the other hand there are strong astrophysical and cosmological evidences about its existence. The most strong evidences come from anisotropy of Cosmic Microwave Background Radiation (CMB), galaxy clusters and galactic rotation curves. A comprehensive chronological history of DM can be obtained from the \cite{hist}. Although its existence remains there evidently, we have poor understanding about its characteristics. There is no alternative candidate for dark matter in Standard Model (SM) of Particle Physics. However, beyond SM, there are some models propose some candidate for DM. MACHOs and primordial black holes, hypothetical structures formed in the early stages of universe, sterile neutrinos, except from active neutrinos, are able to produce thermal relic density and structure formation, axions, which obey Peccei-Quinn mechanism, and Weakly Interactive Massive Particles (WIMPs) can be considered as candidates for DM. Among these candidates WIMPs are the most common alternative candidate for DM. There are a plenty of models proposes DM candidate as WIMP, such as neutralino in RPV SUSY, sneutrinos in MSSM, gravitino in MSUGRA, Z' in 2HDM, lightest Kaluza Klein particle in UED. On the other hand, interactions between DM and SM particles can be via mediator (Simplified Model) or with an effective vertex. In the latter case, the interaction between SM and DM occurs in a hidden sector. Basically in this effective models a cut$-$off scale $\Lambda$ takes the place of a heavy mediator particle with mass $M_*$. Yet, the validity of effective theory of dark matter is another subject of interest. The most general condition for validity of DM EFT is based on the relation between transferred momentum ($Q_tr$) and cut$-$off scale $\Lambda$.\linebreak 
In this paper we always take the validity condition into account, as well. This paper organised as follows: We will give basic assumptions and requirements for fermionic DM of Effective Field Theory in Sect \ref{sec:eft}. In Sect. \ref{sec:direct} the results of DM-nucleus cross section results for direct detection and DM-annihilation velocity averaged cross section to SM particles from galactic halo of a dwarf galaxies will be compared with current experimental data. Then we give a brief summary for all results in Sect. \ref{sec:conc}.

\section{Fermionic Dark Matter of Effective Field Theory}\label{sec:eft}

\begin{figure}[htp!]
    \centering
    \includegraphics[scale=0.5]{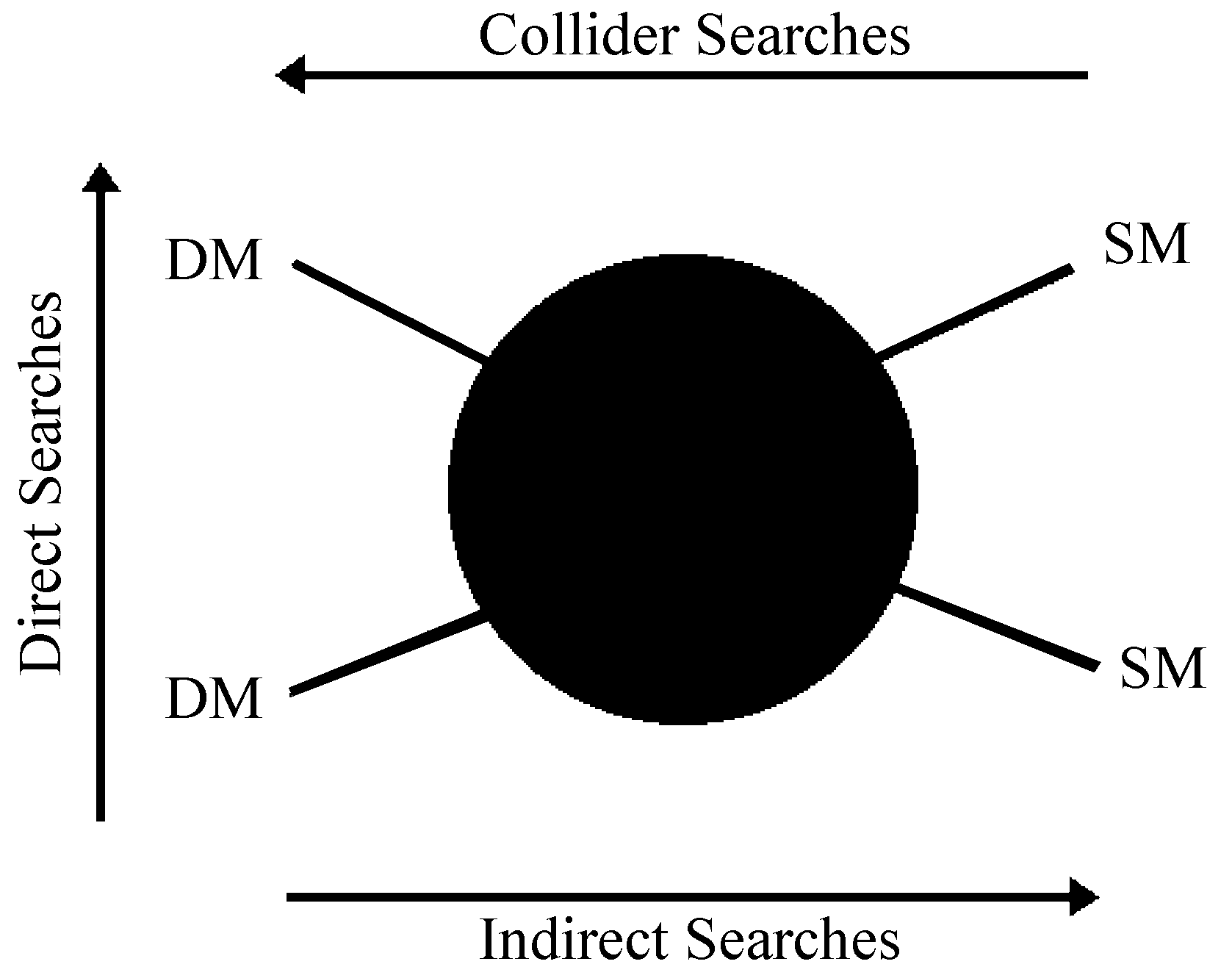}
    \caption{Shematic representation of DM searches}
    \label{fig:dmsearches}
\end{figure}

 Fig. \ref{fig:dmsearches} represents a schema about DM searches. DM annihilation to SM particles is analysed at indirect searches, DM-nucleus interactions is analysed at direct detection, and DM-pair production is analysed at collider searches. There are various techniques for direct detection, indirect detection and collider searches. A comprehensive list of direct detection experiments is presented in Ref. \cite{schumann}, and indirect detection experiments presented in Ref. \cite{gaskins}. These experimental methods fronts diverse DM detection sources, and, therefore diverse DM interactions. Therefore each of these detection techniques leads different kinds of constraints of parameters and limitations of DM model. \\
There are plenty of DM models which have different types of parameters. Effective Field Theory approach is an effortless way to examine a DM model. In order to study EFT, the interactions between DM and SM particles must be defined in a hidden sector via an effective vertex that can be replaced with a heavy mediator particle. A proper parameter space for an effective model must be selected taking care of  direct detection, indirect detection, thermal relic density abundance of DM. \\

For 6-dimensional EFT, fermionic DM is a SM gauge singlet and odd under global $\mathcal{Z}_2$ symmetry. The tree level interactions between this fermionic DM and SM particles can be chosen as the Lagrangian form of the operators given in \cite{ref}.

\begin{itemize}
\item 4-fermion vectoral interactions:
\begin{eqnarray}
		\mathcal{L}_{(uR,dR,eR)\chi}=\frac{g^u_{R}}{2 \Lambda^2}(\bar{u}\gamma^{\mu}u)(\bar{\chi}\gamma_{\mu}\chi) +\frac{g^d_{R}}{2 \Lambda^2}(\bar{d}\gamma^{\mu}d)(\bar{\chi}\gamma_{\mu}\chi) + \frac{g^e_{R}}{2\Lambda^2}(\bar{e}\gamma^{\mu}e)(\bar{\chi}\gamma_{\mu}\chi)  \label{eq-lag-4ferm-vekt}
		\end{eqnarray}
		
\item	4-fermion scalar interactions:
\begin{eqnarray}
		\mathcal{L}_{(\ell,q)\chi}=\frac{g^{\ell}_L}{\Lambda^2}(\bar{\ell}\chi)(\bar{\chi}\ell)  + \frac{g^q_L}{\Lambda^2} (\bar{q}\chi)(\bar{\chi}q) \label{eq-lag-4ferm-sca}
		\end{eqnarray}

\item Fermion-vector-scalar interactions:
\begin{eqnarray}
		\mathcal{L}_{\phi\chi}=\frac{i \alpha_{\phi \chi}}{\Lambda^2}({\phi}^{\dagger}D^{\mu}\phi)(\bar{\chi}\gamma_{\mu}\chi)+h.c.
		\label{eq-lag-ferm-vec-sca}
		\end{eqnarray}
		\end{itemize}

where, $\chi$ is fermionic DM field, $u,d,e$'s are right-handed fermions, $\gamma$'s are gamma matrices, $q,\ell$ denotes left-handed quarks and leptons, $\phi$ is Higgs field  $\Lambda$ is cut-off scale of new physics,  $g_{R}^{u(d,e)}$ and $g_L^{\ell(q)}$'s are the coupling parameters related to dark operators $\alpha$'s. The apparent relation between $g$'s and $\alpha$'s are given as:
		
\begin{eqnarray}
g^{u}_{L}=-\frac{1}{2}\alpha_{q\chi}, \qquad g^{u}_{R}=\frac{1}{2}\alpha_{u\chi} \nonumber
\end{eqnarray}
\begin{eqnarray}
g^{d}_{L}=-\frac{1}{2}\alpha_{q \chi}, \qquad g^{d}_{R}=\frac{1}{2}\alpha_{d\chi} \nonumber
\end{eqnarray}
\begin{eqnarray}
g^{e}_{L}=-\frac{1}{2}\alpha_{\ell \chi},\qquad g^{e}_{R}=\frac{1}{2}\alpha_{e\chi} \nonumber
\end{eqnarray}
\begin{eqnarray}
g^{\nu}_{L}=-\frac{1}{2}\alpha_{\ell \chi}, \qquad g^{\nu}_{R}=0 \nonumber
\end{eqnarray}

\begin{figure}[htp!]
    \centering
    \includegraphics[scale=0.4]{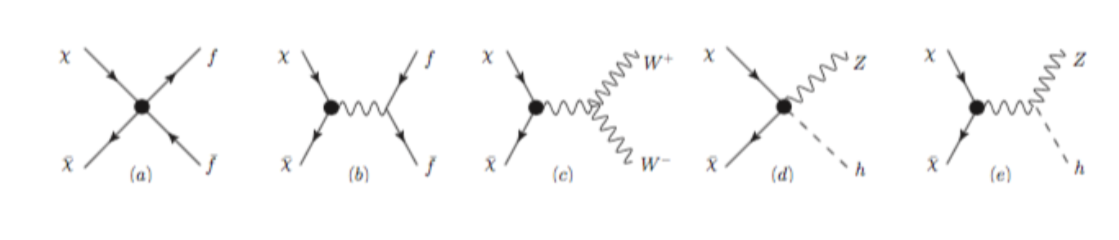}
    \caption{Feynman diagrams of the most dominant processes contributes DM annihilation}
    \label{fig:feyn-ferm}
\end{figure}

The most dominant Feynman diagrams contribute to annihilation of fermionic DM of relic density is showed in Fig.\ref{fig:feyn-ferm}

Blobs in Fig.\ref{fig:feyn-ferm} represent the effective vertices for the interactions. 

\section{Direct and Indirect Detection of Fermionic Dark Matter in Effective Field Theory} \label{sec:direct}

\begin{figure}[htp!]
    \includegraphics[scale=0.5]{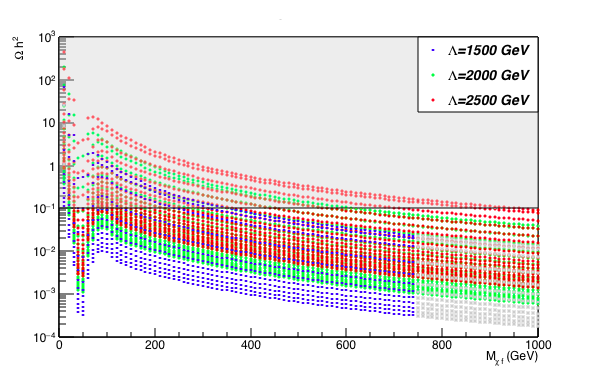}
    \caption{Thermal relic density of fermionic DM in EFT according to model parameters. (Kuday, Ozok, Saka Ref. \cite{kos})}
    \label{fig:relic}
\end{figure}

In order to analyse current experimental status of fermionic dark matter in EFT, one can consider thermal relic density of DM. In the work of Kuday, Ozok and Saka \cite{kos} a comprehensive study for thermal relic density of fermionic DM in EFT is analysed. Model parameters, cut-off scale and mass of DM are constrained by the current value of thermal relic density \cite{planck, wmap}. For this purpose, we developed an effective model file in FeynRules, and implemented this UFO file into MadDM numerical tool to compute dark matter relic density, dark matter nucleus scattering rates and dark matter indirect detection parameters. All information about the model given in Eq. \ref{eq-lag-4ferm-vekt}, Eq. \ref{eq-lag-4ferm-sca}  and Eq. \ref{eq-lag-ferm-vec-sca} inserted to FeynRules, and after checking conservation of all quantum numbers, hermicity of Lagrangians we managed to produce a clean UFO model file. We used this UFO model file to make a comprehensive study of fermionic dark matter. An extensive restrictions of model parameters of fermionic DM in EFT originated from thermal relic density is given in Fig. \ref{fig:relic}.

In Fig. \ref{fig:relic} gray shaded regions are the excluded regions according to upper limit of current thermal relic density, gray data points are excluded according to resulting validity conditions for transfer momentum $\Lambda \geq 2 M_{\chi f}$. Each point denotes $\alpha_{\phi}=1,....,10$ in a given mass and cut-off scale. When $\Lambda$ increases, model exceeds the upper limit of DM thermal relic density. For $\alpha_{u \chi}=\alpha_{d \chi}=\alpha_{e \chi}=\alpha_{q \chi}=\alpha_{l \chi}=\alpha_{\phi \chi}=1$, the proper parameter space of model:

\begin{itemize}
\item For $\Lambda=1.5$ TeV, mass of fermionic DM $M_{\chi f}\geq 250$ GeV, 
\item For $\Lambda=2$ TeV, mass of fermionic DM $M_{\chi f}\geq 460$ GeV,
\item For $\Lambda=2.5$ TeV, mass of fermionic DM $M_{\chi f}\geq 720$ GeV,
\end{itemize}

Another DM search domain is direct detection of dark matter via elastic or inelastic scattering of dark matter candidate from a nucleus. There are a number of experiments hunts for DM candidate through direct detection like experiments XENON1T \cite{xenon} and LUX \cite{lux} and with $^{19}F$ targetted PICO60 \cite{pico} experiments. By using  \textsc{MadDM v.3.0} direct detection module, the resulting DM-nucleus scattering spin dependent (SD) and spin independent (SI) scattering cross sections for 6-dimensional EFT of fermionic DM are depicted in Fig. \ref{fig:mxf-direct}:

\begin{figure}[htp!]
    \centering
    \includegraphics[scale=1.2]{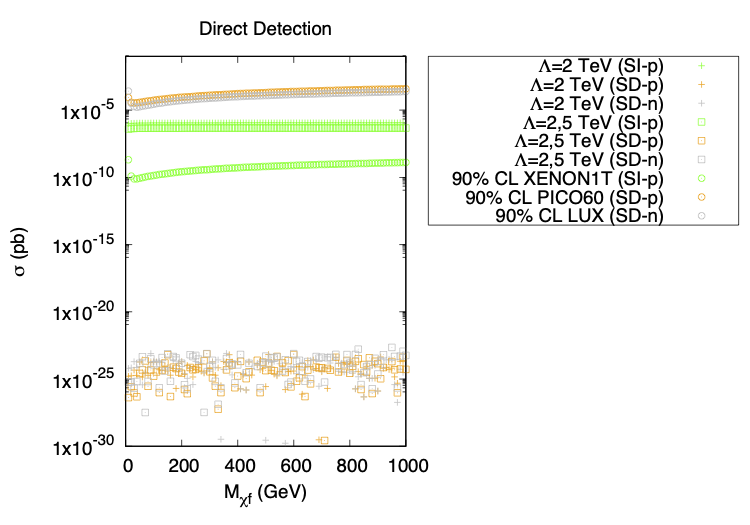}
    \caption{For $\Lambda=2$ TeV and $\Lambda=2.5$ TeV, obtained results of DM-nucleus scattering cross sections and experimental upper limits for cross sections in 6-Dimensional EFT of fermionic DM.}
    \label{fig:mxf-direct}
\end{figure}

In Fig. \ref{fig:mxf-direct} how the scattering cross sections of DM-nucleus change due to the mass of fermionic DM candidate in 6-dimensional EFT is depicted. Data points with $+$ and $\square$ represent the results for $\Lambda=2$ TeV and $\Lambda=2.5$ TeV for EFT. Data points with $\odot$ represent the experimental upper limits of direct detection experiments with $\%90$ CL. Green points represent spin independent neutron scattering cross sections (experimentally, cross sections of XENON1T), orange points represent spin dependent elastic proton scattering cross sections (experimentally, cross sections of PICO60) and gray points represent spin dependent elastic neutron scattering cross section (experimentally, cross sections of LUX). As a result, obtained results of effective theory of fermionic DM, are excluded by XENON1T experiments and, nevertheless results are still acceptable by LUX and PICO60 experiments.  \\

\begin{figure}[htp!]
    \centering
            \includegraphics[scale=0.8]{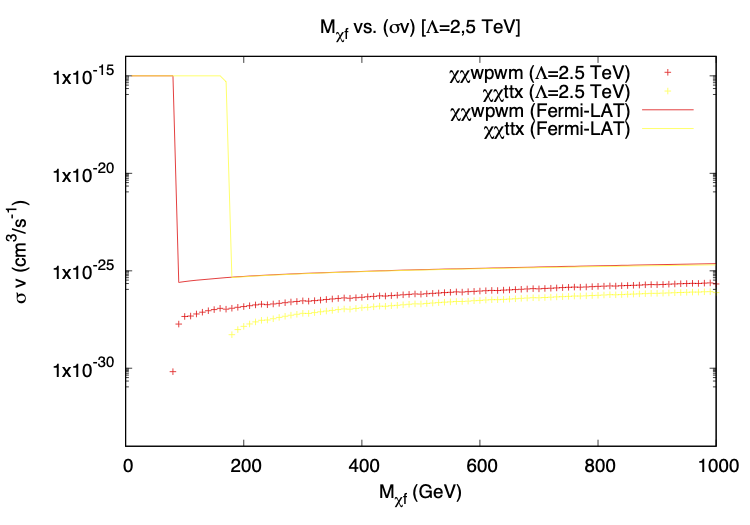}
    \caption{The comparison of velocity averaged cross sections of fermionic DM pair annihilation into SM particles with Fermi-LAT data(6 years observations)}
    \label{fig:ferm-indirect}
\end{figure}

Another survey area of DM is indirect detection of DM, which is based on DM annihilation. Indirect detection experiments basically investigates two DM annihilation into SM particles or gamma rays. There are a plenty of indirect detection experiments based on large area telescopes, ground-based atmospheric Cherenkov imaging telescopes and other imaging techniques. In this paper we also investigated fermionic DM pair annihilation into SM particles (indirect detection), by using \textsc{MadDM v.3.0} indirect detection module. The velocity averaged cross sections of DM pair annihilation into possible SM particles are shown in Fig. \ref{fig:ferm-indirect}:

\begin{figure}[htp!]
    \centering
    \includegraphics[scale=0.8]{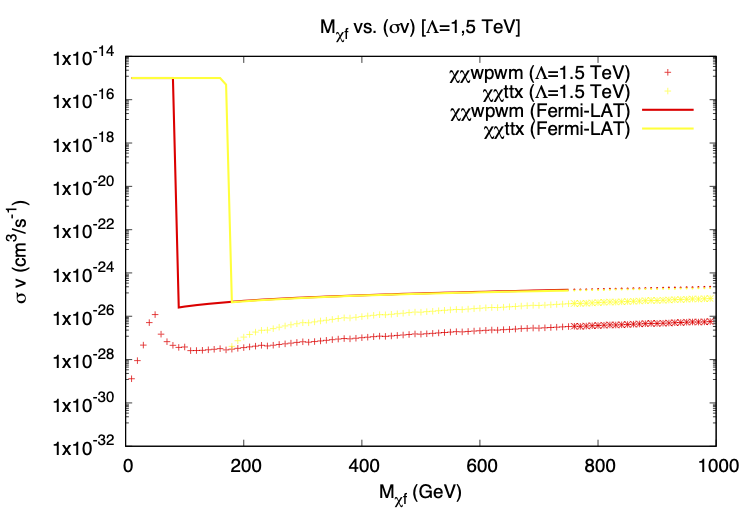}
\end{figure}
\begin{figure}[htp!]
    \centering
        \includegraphics[scale=0.8]{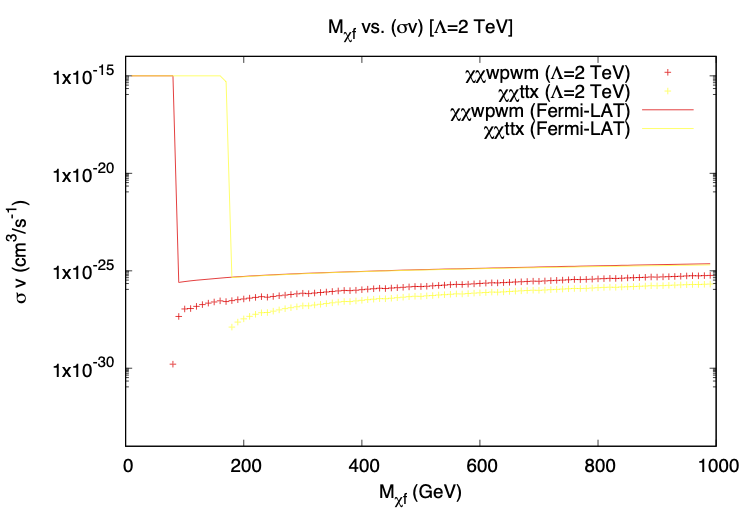}
\end{figure}

In Fig. \ref{fig:ferm-indirect} for  $\Lambda=1.5$ TeV, $\Lambda=2$ TeV and $\Lambda=2.5$ TeV cut-off of the effective theory, the fermionic DM annihilation into SM particles velocity averaged cross sections  $\sigma v$ are compared with 6 years observed data of Fermi-LAT dSphs (\%90 CL). The dashed line represents the Fermi-LAT results \cite{fermilat}, the straight line represents the obtained results of effective theory of fermionic DM. DM pair annihilation into leptons and light fermions cross sections to  are too small to be shown in the figure. Nevertheless, the velocity averaged cross sections are in acceptable region according to upper limits of corresponding leptons and light fermion of Fermi-LAT data. Moreover, the highest value for cross section of fermionic DM pair annihilation comes from $\chi \bar{\chi} W^+W^-$, $\chi \bar{\chi} t \bar{t}$ decay channels. From Fig. \ref{fig:ferm-indirect}, it can be easily noticed that the obtained results for 6-dimensional EFT for fermionic DM pair annihilation into dominant SM particles are compatible with Fermi-LAT observed data.

\section{Summary and Conclusion}\label{sec:conc}
In this study, we analysed fermionic dark matter of 6-dimensional effective field theory at direct and indirect research areas of dark matter searches. We set a specified tools in order to analyze for fermionic dark matter in the model of 6-dimensional EFT. First, the relic density of fermionic dark matter of six dimensional EFT calculated by scanning all parameter space of the model. Afterwards the proper parameter region is determined  according to the current result of dark matter relic density of Planck results \cite{planck}. Spin-dependent and spin-independent scattering cross section of fermionic DM from nucleus of various direct detection experiments are presented in Fig. \ref{fig:mxf-direct}. For values of cut-off scale of $\Lambda=2$ TeV and $\Lambda=2.5$ TeV the obtained scattering cross-sections results of the effective model from the experiments, are compatible within 90 \% CL upper limits of PICO60 and LUX experiment results. In Fig. \ref{fig:ferm-indirect} the velocity averaged cross section of DM annihilation into SM particles in Dwarf spheroidal galaxies (dSPhs) are compared with the results of six-years data of Fermi-LAT. The velocity averaged cross section of DM annihilation into leptonic particles are too tiny to place to graph. Yet, the leptonic annihilations are get along with the upper limits of velocity averaged cross sections for leptonic annihilations of Fermi-LAT data. Notable annihilation channels are demonstrated in the Fig. \ref{fig:ferm-indirect}. It can be seen from the figure that DM annihilation to all possible SM particles are in the proper upper limit of Fermi-LAT data. Consequently, the model file we set in six-dimensions present a proper framework  for all type of search platform for DM research. The obtained results for nucleus scattering cross sections with \%90 CL are compatible with  of direct detection experiments (PICO60 and LUX) and and the obtained results for velocity averaged annihilation cross section to SM particles are consistent indirect detection.


\begin{acknowledgments}
In this work, Ayşe Elçiboğa Kuday was supported by TUBITAK 2211/C Priority Areas of Domestic PhD Scholars.
 
\end{acknowledgments}


\end{document}